\documentstyle[emulateapj]{article}

\begin{document}
\title{Hierarchical Structure Formation and
Chemical Evolution of Damped Ly$\alpha$ Systems} 
\author{Y.-Z. Qian\altaffilmark{1} and G.  J.  Wasserburg\altaffilmark{2}}
\altaffiltext{1}{School of Physics and Astronomy, University of Minnesota,
Minneapolis, MN 55455; qian@physics.umn.edu.}  
\altaffiltext{2}{The Lunatic Asylum, Division of Geophysics and Planetary 
Sciences, California Institute of Technology, Pasadena, CA 91125.}  


\begin{abstract}
We present a model for chemical evolution of damped Ly$\alpha$
systems considering production of ``metals'' by SNe II
and infall associated with hierarchical structure formation.
The growth of metallicity in these systems is a
reflection of the competition between astration
and infall. The apparent late turn-on of these
systems is due to the late cut-off of infall. The wide range in
[Fe/H] at a given redshift is explained by the range 
of the times for onset of star formation and the range of the times 
for infall cessation in different systems. 
The observed lower bound of
[Fe/H]~$\approx -3$ follows from
the very rapid initial rise of [Fe/H] subsequent to onset of star
formation. To reach [Fe/H]~$\approx -3$ from a metal-free
initial state requires only $\sim 30$ Myr so that the probability
of observing lower [Fe/H] values is very small.
\end{abstract}


\section{Introduction}
We discuss chemical evolution of damped Ly$\alpha$ 
(DLA) systems based on hierarchical structure formation. Observations
(Lu et al. 1996; Prochaska \& Wolfe 2000, 2002; Prochaska et al. 2003) 
show there is a baseline enrichment of
[Fe/H]~$=\log({\rm Fe/H})-\log({\rm Fe/H})_\odot\approx -3$
for DLA systems. Of 96 systems over the redshift range $0.5<z<5$, 
the lowest observed [Fe/H] is $-3.13$ although the newer observations 
could have detected much lower values (Prochaska et al. 2003). Furthermore, 
there is a large dispersion in [Fe/H] at any fixed $z$.

In earlier discussion (Wasserburg \& Qian 2000b), we treated DLA
systems as closed systems formed from the intergalactic medium
(IGM) with a fixed initial inventory of ``metals'' (the prompt inventory)
corresponding to [Fe/H]$_P\approx -3$ and assumed that subsequent Fe
enrichment of each system was governed by a constant Fe production rate
per H atom in the gas of $P_{\rm Fe}/({\rm H})$. 
Element production started when a system was formed at time $t^*$ after 
big bang. At time $t>t^*$, the number ratio (Fe/H) of Fe atoms to H atoms
in the gas evolves according to
\begin{equation}
\frac{d({\rm Fe/H})}{dt}=\frac{P_{\rm Fe}}{\rm (H)}.
\label{feh}
\end{equation}
We only treat Fe contributions from Type II supernovae
(SNe II) in all our considerations as SNe Ia turn on at later times.
Integration of equation (\ref{feh}) gives
\begin{equation}
({\rm Fe/H})=({\rm Fe/H})_P
+\frac{P_{\rm Fe}}{\rm (H)}(t-t^*).
\label{feht}
\end{equation}
As SNe II contributed
$\sim {\rm (Fe/H)}_\odot/3$ over the period
of $(t-t^*)\sim 10$ Gyr prior to solar system formation,
$P_{\rm Fe}/({\rm H})$ is estimated as
$\sim {\rm (Fe/H)}_\odot/(30\ {\rm Gyr})$.

In this model, the baseline Fe enrichment is explained by the prompt 
inventory and the dispersion in [Fe/H] at a fixed $z$ by 
the range of $t^*$. It was found that most
systems at any given $z$ have $t^*$ close to the age $t(z)$ of the
universe at this $z$. This implies that the ``turn-on'' of DLA
systems occurs rather long after the big bang.
Further, essentially all the data lie below the upper bound for 
[Fe/H] corresponding to $t^*=0$.
The above model appears to provide a reasonable description of
the data on [Fe/H] for DLA systems.
This is rather remarkable considering that the model ignores infall.
According to 
hierarchical structure formation, infall is
essential to formation of DLA systems. Here we examine chemical
evolution of these systems by including infall. 
We wish to gain some insights
into what causes most DLA systems at a given $z$ to have
$t^*$ close to $t(z)$ in the closed-system model. We also address whether 
the prompt inventory is needed to explain the baseline Fe 
enrichment for DLA systems.

\section{Chemical Evolution with Infall}
Consider a system of gas and stars with infall of primordial metal-free
gas. The equations for evolution of the numbers of Fe and H atoms 
[(Fe) and (H)] in the gas are
\begin{eqnarray}
\frac{d({\rm Fe})}{dt}&=&P_{\rm Fe}+({\rm Fe/H})
\frac{d({\rm H})_{\rm as}}{dt},\label{fei}\\
\frac{d({\rm H})}{dt}&=&\frac{d({\rm H})_{\rm as}}{dt}+
\frac{d({\rm H})_{\rm in}}{dt},\label{hi}
\end{eqnarray}
where $P_{\rm Fe}$ is the Fe production rate of SNe II,
$d({\rm H})_{\rm as}/dt<0$ is the astration rate, and
$d({\rm H})_{\rm in}/dt>0$ is the infall rate. When
$|d({\rm H})_{\rm as}/dt|$ is small compared to $d({\rm H})_{\rm in}/dt$,
(H) is governed by infall. Equations (\ref{fei}) and
(\ref{hi}) give (Qian \& Wasserburg 2003)
\begin{equation}
\frac{d({\rm Fe/H})}{dt}=\frac{P_{\rm Fe}}{\rm (H)}-\frac{1}{\rm (H)}
\frac{d({\rm H})_{\rm in}}{dt}({\rm Fe/H}).
\label{fehi}
\end{equation}
Formal integration of equation (\ref{fehi}) gives
\begin{equation}
({\rm Fe/H})=\frac{P_{\rm Fe}}{\rm (H)}t-\int_0^{t}\frac{1}{\rm (H)}
\frac{d({\rm H})_{\rm in}}{dt'}({\rm Fe/H})dt'.
\label{fehit}
\end{equation}
With the integral term
interpreted as $P_{\rm Fe}t^*/{\rm (H)}$,
this resembles equation (\ref{feht}) for (Fe/H)$_P=0$. 
So the effect of a sustained large 
infall rate is similar to late ``turn-on.''

In general, equations (\ref{hi}) and (\ref{fehi}) must be solved together.
For simplicity, we first consider the case where
$d({\rm H})_{\rm in}/dt\gg |d({\rm H})_{\rm as}/dt|$
so ${\rm (H)}\approx {\rm (H)}_{\rm in}$. We then have
\begin{equation}
\frac{dZ_{\rm Fe}}{dt}=\lambda_{\rm Fe}-\lambda_{\rm in}Z_{\rm Fe},
\label{zfe}
\end{equation}
where $Z_{\rm Fe}\equiv {\rm (Fe/H)/(Fe/H)}_\odot$,
$\lambda_{\rm Fe}\equiv P_{\rm Fe}/[({\rm Fe/H})_\odot{\rm (H)}]$,
and $\lambda_{\rm in}\equiv d\ln({\rm H})_{\rm in}/dt$. 
If $dZ_{\rm Fe}/dt$ is negligible, a quasi-steady state is achieved and
$Z_{\rm Fe}(t)$ approximately assumes the value
$Z_{\rm Fe}^{\rm QSS}(t)\equiv\lambda_{\rm Fe}(t)/\lambda_{\rm in}(t)$.
With onset of Fe production at $t_0$, we have
\begin{equation}
Z_{\rm Fe}(t)\approx Z_{\rm Fe}^{\rm QSS}(t)-Z_{\rm Fe}^{\rm QSS}(t_0)
\exp\left[-\int_{t_0}^t\lambda_{\rm in}(t')dt'\right]
\label{zfeta}
\end{equation}
for $\lambda_{\rm in}(t)\gg |d\ln Z_{\rm Fe}^{\rm QSS}/dt|$.
At $t>t_0+[\lambda_{\rm in}(t_0)]^{-1}$, the exponential term 
is negligible and 
$Z_{\rm Fe}(t)\approx Z_{\rm Fe}^{\rm QSS}(t)$.

We consider a baryonic system
that is formed through infall into the potential
well of a dark matter halo. The mass
$M$ of the halo grows according to hierarchical structure formation. 
We assume that baryonic and
dark matter are fed into the halo at a fixed mass ratio. 
The infall rate is then
\begin{equation}
\lambda_{\rm in}(t)=\frac{d\ln M}{dt}=\frac{d\ln M}{dz}\frac{dz}{dt}.
\label{inf}
\end{equation}
With $t\approx 17(1+z)^{-3/2}$ Gyr at $z>0.5$, equation (\ref{inf}) gives
\begin{equation}
\lambda_{\rm in}(t)\approx\frac{4.4}{\rm Gyr}
\left(\frac{\rm Gyr}{t}\right)^{5/3}
\left|\frac{d\ln M}{dz}\right|.
\label{inft}
\end{equation}

From Figure 6 in Barkana \& Loeb (2001), we find that 
$|d\ln M/dz|\sim 2.3$ (within a factor of 2) for $0.5<z<5$.
As an example, we take $\lambda_{\rm in}(t)=(0.1\ {\rm Gyr})^{-1}
({\rm Gyr}/t)^{5/3}$ and $\lambda_{\rm Fe}=(30\ {\rm Gyr})^{-1}$
for $t>0$ (i.e., $t_0=0$). We numerically integrate
equation (\ref{zfe}) and show the
evolution of [Fe/H]~$=\log Z_{\rm Fe}$ in Figure 1. The quasi-steady 
state value
$Z_{\rm Fe}^{\rm QSS}(t)=\lambda_{\rm Fe}/\lambda_{\rm in}(t)=
(t/{\rm Gyr})^{5/3}/300$
is a good approximation to the exact solution. 
For the case of evolution from a metal-free initial
state without infall, $Z_{\rm Fe}(t)=\lambda_{\rm Fe}t$ 
(the case of $t^*=0$ in \S1). 
In comparison with this, Fe enrichment 
in the case of infall is suppressed by a factor of
$\approx\lambda_{\rm in}(t)t\approx 10({\rm Gyr}/t)^{2/3}$, which ranges 
from 4.5 for $t=3.3$ Gyr ($z=2$) to 8.9 for $t=1.2$ Gyr ($z=5$).
This would explain the late ``turn-on'' requiring $t^*$ to be close to
$t(z)$ in the closed-system model. The data on [Fe/H] for 96 DLA systems 
(Prochaska et al. 2003) are shown in Figure 1.
It can be seen that the solid curve for the case of infall passes
through the median of the body of the data. This curve
has a slope of $d{\rm [Fe/H]}/dz\approx -1.1/(1+z)$. For $z\gtrsim 2$
where most of the data lie, this slope is in good agreement with
the estimate of $d{\rm [Fe/H]}/dz\approx -0.26$ given by
Prochaska et al. (2003) for the growth of the mean [Fe/H] as
a function of $z$. However, the 
above result does not explain the wide range in [Fe/H] at a given $z$.

\section{The Dispersion in [Fe/H]}
We now turn to the wide range in [Fe/H] for DLA 
systems at a given $z$. We assume a constant Fe production
rate $\lambda_{\rm Fe}$ per H atom in the gas for $t>t_0$.
The infall rate $\lambda_{\rm in}$ is estimated based on
hierarchical structure formation and we now allow the possibility
that infall may cease at a time $t_{\rm IC}$.
We assume that the infall rate greatly exceeds the astration rate
at $t<t_{\rm IC}$. Under these
assumptions, $t_0$ and $t_{\rm IC}$ 
determine [Fe/H] in a baryonic system at time $t$
(see eqs. [\ref{zfeta}] and [\ref{zfetb}]).

First consider the case where $t_0=0$ and infall 
ceases at $t_{\rm IC}$. The evolution of $Z_{\rm Fe}$
is the same as for the case of infall discussed in \S2 until
$t=t_{\rm IC}$. At $t>t_{\rm IC}$, equation (\ref{zfe}) reduces to
$dZ_{\rm Fe}/dt=\lambda_{\rm Fe}$, which gives
\begin{equation}
Z_{\rm Fe}(t)=Z_{\rm Fe}(t_{\rm IC})+\lambda_{\rm Fe}(t-t_{\rm IC}).
\label{zfetb}
\end{equation}
Solutions for $t_{\rm IC}=0$, 1.5 Gyr ($z_{\rm IC}=4$), and 
4.3 Gyr ($z_{\rm IC}=1.5$) are shown in Figure 2.
In all cases of $t_{\rm IC}>0$, the evolution of [Fe/H] for 
$t<t_{\rm IC}$ is along the solid curve labeled ``continuous infall.''
As illustrated by points A, B, and C at $z=3$ in Figure 2, a wide range 
of [Fe/H] bounded by the cases of no infall $(t_{\rm IC}=0)$ and 
continuous infall can be produced at a given $z$ for different 
$t_{\rm IC}$ values.

Next consider the case where infall is continuous but $t_0$ may vary.  
Subsequent to onset of Fe production at $t_0$,
$Z_{\rm Fe}$ grows rapidly and approaches $Z_{\rm Fe}^{\rm QSS}$
on a timescale of $\sim\lambda_{\rm in}(t_0)^{-1}$ (see eq. 
[\ref{zfeta}]). This is shown in Figure 1 for
$t_0=1.5$ Gyr ($z_0=4$), 2.1 Gyr ($z_0=3$), and 3.3 Gyr ($z_0=2$).
For example, [Fe/H] in baryonic systems formed
at $t_0=2.1$ Gyr first evolves rapidly, crossing
[Fe/H]=-3, and then grows more slowly
toward the quasi-steay state solution.
Thus, values of [Fe/H] below those for the case of $t_0=0$
may be populated with baryonic systems that have different $t_0$ values.
Then the lower range in [Fe/H] at 
a given $z$ can also be explained by the infall model. 

The observed lower bound of [Fe/H]~$\approx -3$ for DLA
systems requires attention. With infall, the growth of $Z_{\rm Fe}$ is
determined by competition between $\lambda_{\rm Fe}$ and 
$\lambda_{\rm in}Z_{\rm Fe}$. As $Z_{\rm Fe}(t_0)=0$, 
$\lambda_{\rm Fe}$ governs the initial growth
of $Z_{\rm Fe}$ so long as $\lambda_{\rm Fe}/\lambda_{\rm in}\gg 
Z_{\rm Fe}$. We note that 
$\lambda_{\rm Fe}/\lambda_{\rm in}=(t/{\rm Gyr})^{5/3}/300\gg 10^{-3}$
for $t\gg 0.49$ Gyr ($z\ll 9.6$). So for systems with $t_0>1.2$ Gyr 
($z_0<5$), the growth of $Z_{\rm Fe}$ up to $10^{-3}$ is essentially
governed by $\lambda_{\rm Fe}=(30\ {\rm Gyr})^{-1}$. These
systems achieve $Z_{\rm Fe}=10^{-3}$
in $\sim 30$ Myr and then approach
$Z_{\rm Fe}\approx Z_{\rm Fe}^{\rm QSS}>10^{-3}$ on a timescale of
$\sim\lambda_{\rm in}(t_0)^{-1}$ (see eq. [\ref{zfeta}] and Fig. 1). 
Thus, all baryonic systems will reach $Z_{\rm Fe}=10^{-3}$ or greater
in a very short time at $z<5$ and the probability for observing 
DLA systems with [Fe/H]~$<-3$ is quite small.

The ranges of $t_0$ and $t_{\rm IC}$ in the infall model
to explain the dispersion in [Fe/H] at a given $z$ are analogous to 
the range of $t^*$ in the closed-system model. 
However, uniform growth of [Fe/H] occurs only at
$t>t_{\rm IC}$ in the infall model. It is the dilution of 
gas by fresh infalling baryonic material that retards [Fe/H]
from uniform growth although Fe production and astration 
has been going on since $t_0$.
To relate $t_0$ and $t_{\rm IC}$ for
a baryonic system to properties of the
halo hosting the system, we follow the discussion 
in Barkana \& Loeb (2001). The condition for a halo of mass $M$
associated with an $n\sigma$ density fluctuation (an $n\sigma$ halo) 
to collapse at redshift $z$ is
\begin{equation}
1.33(1+z)\approx n\sigma(M),
\label{zn}
\end{equation}
where $\sigma(M)$ is shown in Figure 5 of Barkana \& Loeb 
(2001). Equation (\ref{zn}) gives the $z$ values at which a fixed mass 
$M$ is reached by halos with different $n$ values.

A $1\sigma$ halo today has $M\approx 10^{13}\,M_{\odot}$ and should 
contain $\approx 10^{12}\,M_{\odot}$
of baryonic matter. This mass is far 
greater than found for a typical galaxy. Thus, we may consider that 
infall into individual protogalaxies ceases
when the accreted baryonic matter reaches 
$\sim 10^{10}$--$10^{11}\,M_{\odot}$ in a halo
of $\sim 10^{11}$--$10^{12}\,M_{\odot}$.
In the hierarchical model, a halo of larger
masses (e.g., $\gtrsim 10^{12}\,M_{\odot}$) may consist of a number of
smaller collapsed halos instead of a single object. This also applies 
to the associated baryonic systems. We assume that infall ceases 
at $t_{\rm IC}$ when 
a halo reaches a mass of $M_{\rm IC}=10^{11}\,M_{\odot}$ 
[$\sigma(M_{\rm IC})=3.5$]. This occurs at redshifts $z_{\rm IC}=4$
and 1.5 for $1.9\sigma$ and $0.95\sigma$ halos, respectively.
These redshifts correspond to the
$t_{\rm IC}$ in Figure 2 and seem reasonable.
For $>4\sigma$ halos, $M_{\rm IC}=10^{11}\,M_{\odot}$ is reached at
$z_{\rm IC}>9.5$ ($t_{\rm IC}<0.5$ Gyr). The evolution
of [Fe/H] at $z<5$ ($t>1.2$ Gyr) for baryonic systems inside these 
high $\sigma$ halos  
is close to the case of $t_{\rm IC}=0$.
Similar results are found for $M_{\rm IC}\sim 
10^{11}$--$10^{12}\,M_{\odot}$.

Now consider $t_0$, which may be 
taken as the onset of star formation. Following the
discussion in Barkana \& Loeb (2001), the first stars appear to have
formed in $\sim 3\sigma$ to $4\sigma$ halos of mass 
$M\sim 10^5\,M_\odot$ [$\sigma(M)\approx 10$]
at $z\sim 20$--30. For these halos, $t_0$ is only
$\sim 0.1$ Gyr. Formation of stars at $z<20$ appears to require a
minimum halo mass of $\sim 10^8$--$10^9\,M_\odot$.
We assume that star formation starts at $t_0$ when a halo reaches a
mass of $M_0=5\times 10^8\,M_\odot$ [$\sigma(M_0)=6$]. This occurs
at redshifts $z_0=4$, 3, and 2 for $1.1\sigma$, $0.89\sigma$, and 
$0.67\sigma$ halos, respectively. These redshifts correspond to
the $t_0$ in Figure 1 and again seem
reasonable. Similar results are found for $M_0\sim 
10^8$--$10^9\,M_{\odot}$.

It follows that at a given $z$, baryonic
systems inside different halos are in different stages of evolution.
Consider an example using $M_0=5\times 10^8\,M_\odot$ and
$M_{\rm IC}=10^{11}\,M_{\odot}$. At $z=2.6$, star formation just
starts in a $0.8\sigma$ halo ([Fe/H]~$\sim -3$), 
Fe production competes with infall
in a $1.1\sigma$ halo ([Fe/H]~$=-1.9$, see Fig. 1), 
and uniform growth of [Fe/H] has been going on since infall
cessation at $z_{\rm IC}=4$ in a $1.9\sigma$ halo ([Fe/H]~$=-1.4$,
see Fig. 2). For a sample of DLA systems at $z=2.6$,
systems with $-3<[{\rm Fe/H}]\leq -1.9$,
$-1.9<[{\rm Fe/H}]\leq -1.4$, and
$[{\rm Fe/H}]>-1.4$ are then associated with
$n\sigma$ halos with $0.8<n\leq 1.1$,
$1.1<n\leq 1.9$, and $n>1.9$, respectively.
Statistically speaking, the fraction
of halos more evolved than an $n\sigma$ halo is
$F(n)=\sqrt{2/\pi}\int_n^\infty\exp(-x^2/2)dx$. Thus, we expect
the occurrences of $-3<{\rm [Fe/H]}\leq -1.9$, 
$-1.9<{\rm [Fe/H]}\leq -1.4$,
and [Fe/H]~$>-1.4$ at $z=2.6$ to be in the ratios of 
$[F(0.8)-F(1.1)]:[F(1.1)-F(1.9)]:F(1.9)\approx 0.7:1:0.3$. Of
19 DLA systems at $2.4<z<2.7$, the numbers of systems in the above 
[Fe/H] intervals are 5, 10, and 4, consistent with expectation from 
our model.

\section{Discussion and Conclusions}
We have treated chemical evolution of DLA systems considering 
astration and infall. It is assumed
that the H in the gas is initially controlled by 
infall of primordial metal-free baryonic matter and the Fe production 
rate per H atom in the gas is constant subsequent to onset of star 
formation in a system. With the infall rate estimated from the standard 
scenario of hierarchical structure formation,
this model yields an explanation for the data on [Fe/H] in DLA systems.
It is shown that the slow growth of [Fe/H] with decreasing $z$ is a direct
consequence of competition between enrichment by Fe production 
and dilution by infall. It is argued that the upper range of 
[Fe/H] at a given $z$ results from the different times at which
individual halos associated with protogalaxies
reach a mass of $\sim 10^{11}\,M_\odot$ and infall ceases
or greatly diminishes. It is also
argued that the lower range of [Fe/H] results 
from the different times at which individual halos reach a minimum
mass of $\sim 10^8$--$10^9\,M_\odot$
required to initiate astration.  
Because the initial growth of [Fe/H] up to $-3$ is very rapid
subsequent to onset of astration, the probability of observing
DLA systems with [Fe/H]~$<-3$ is very small. The approach presented here
appears to be a reasonably quantitative description of chemical 
evolution of DLA systems that is compatible with the
paradigm of hierarchical structure formation. Of the available data,
only three points lie outside the upper bound of the model
and these are still within a factor of 2 of this bound.

We assumed the astration rate is small compared to the infall
rate during the infall phase. If we
assume a constant astration rate $\alpha$
per H atom in the gas consistent with the Fe production rate, then
equation (\ref{hi}) reduces to $d({\rm H})/dt=-\alpha({\rm H})
+d({\rm H})_{\rm in}/dt$, which can be solved explicitly. As long as
$d({\rm H})_{\rm in}/dt>\alpha({\rm H})$, (H) increases. When infall
ceases, (H) reaches the maximum value (H)$_{\rm max}$ and the star
formation rate also reaches the maximum value 
$\alpha{\rm (H)}_{\rm max}$. Subsequently, the gas will be depleted
by star formation on a timescale of $\sim\alpha^{-1}$. 
When a protogalaxy would reach its maximum star formation
rate depends on when its associated halo reaches a mass of
$\sim 10^{11}\,M_\odot$ and infall ceases. This occurs at $z\sim 1.6$
and 4.3 for $1\sigma$ and $2\sigma$ halos, respectively. 
The extent to which the approach outlined here can be
used to quantitatively explain the cosmic star formation history
remains to be explored.

In earlier works (Wasserburg \& Qian 2000a; Qian \& Wasserburg 2002),
we proposed that very massive
($\gtrsim 100\,M_\odot$) stars (VMSs) produced the elements from
C to the Fe group to explain the observed jump in
the abundances of heavy $r$-process elements
(Ba and above) at [Fe/H] $\approx -3$. The VMSs produced no heavy 
$r$-elements but dominated
chemical evolution at [Fe/H]~$<-3$. This evolution resulted in
a prompt inventory of ``metals'' corresponding to [Fe/H]~$\approx -3$ 
in the IGM. 
Cessation of VMS 
activities and rapid occurrence of a hypothesized SN II source for 
heavy $r$-elements then led to an explanation for the observed
jump in abundances of these elements at [Fe/H]~$\approx -3$.
The dominance of VMS activities at [Fe/H]~$<-3$ and transition
to normal astration at [Fe/H]~$\approx -3$ are in
accord with the theoretical study of Bromm et al. (2001). They showed 
there is a critical
metallicity corresponding to $-4<{\rm [Fe/H]}<-3$ above which
normal star formation should occur. Oh et al. (2002) showed that
the number of VMSs required to produce [Fe/H]~$\approx -3$ would
also provide enough photons to reionize the IGM. 
Qian, Sargent, 
\& Wasserburg (2002) showed that the abundances
of C, O, and Si in Ly$\alpha$ forests appear to be consistent with
the corresponding prompt inventory. The observed
lower bound of [Fe/H]~$\approx -3$ for DLA systems
was interpreted by Wasserburg \& Qian (2000b) as strong evidence
in support of the prompt inventory model. 
Thus, there was a general
accord between very different approaches.

There are now two considerations that require a re-evaluation of
the prompt inventory model. 
Qian \& Wasserburg (2003) proposed that
the possible sources for heavy $r$-elements are accretion-induced
collapse (AIC) of white dwarfs and low-mass ($\sim 8$--$10\,M_\odot$)
SNe II, with the most likely source being AIC events. If AIC
events instead of SNe II
are the dominant source for heavy $r$-elements, the jump in
abundances of these elements at [Fe/H]~$\approx -3$ can be
attributed to the delay in turn-on of these events relative to SNe II
without invoking VMSs.
We showed here that the growth of [Fe/H] to $-3$ from a metal-free 
initial state is very rapid (in $\sim 30$ Myr).
So the probability of finding DLA systems
with [Fe/H]~$<-3$ is very low and the observed lower bound of
[Fe/H]~$\approx -3$ can no longer be considered
as an argument in favor of the prompt inventory model. 
A prompt inventory corresponding to
[Fe/H]~$\approx -3$ is thus not required to explain the data 
on DLA systems but is
also not in conflict with the interpretation
of these data using the infall model. However,
if there is no prompt inventory, 
a question remains regarding the approximately constant metallicity 
of the IGM over $1.5\leq z\leq 5.5$ (Songaila 2001; 
Pettini et al. 2003). At present, it is not
obvious how this can be explained by the infall model. 
Without VMSs associated with production of the prompt 
inventory, the source for reionization may also be a problem.

\acknowledgments
We wish to dedicate this paper to Allan Sandage
for galaxies of reasons.
We thank Jason Prochaska for making available the new data
that stimulated us to further consider DLA systems.
Intense questioning by Michael Norman on the requirement of VMSs 
and a prompt inventory in our earlier models also provided a great
stimulus. This work was supported in part by DOE grants 
DE-FG02-87ER40328, DE-FG02-00ER41149 (Y. Z. Q.) and DE-FG03-88ER13851 
(G. J. W.), Caltech Division Contribution 8906(1108).


\figcaption{Data (squares) on [Fe/H] for 96 DLA systems 
(Prochaska et al. 2003). The solid curve labeled ``infall'' shows the 
evolution of [Fe/H] for a system with infall and a constant Fe 
production rate $\lambda_{\rm Fe}$ per
H atom in the gas since
the big bang. The thin dashed curve is the quasi-steady
state approximation. 
The solid curve labeled ``no infall'' is for a system
with a constant $\lambda_{\rm Fe}$ since the big bang but with no infall.
The short-dashed curve is for a system in which 
infall starts at the big bang but star formation
starts at $t_0=1.5$ Gyr ($z_0=4$), the dot-dashed curve is for
$t_0=2.1$ Gyr ($z_0=3$), and the long-dashed curve is for $t_0=3.3$ Gyr 
($z_0=2$). Assuming a halo mass of $M_0=5\times 10^8\,M_\odot$
at the onset of star formation, $z_0=4$, 3, and 2 
correspond to $1.1\sigma$, $0.9\sigma$, and $0.7\sigma$ density 
fluctuations, respectively. Due to very rapid initial rise of [Fe/H], 
the probability of sampling DLA 
systems with [Fe/H]~$<-3$ is very small. 
The prompt inventory 
proposed earlier (horizontal dashed line) is shown for reference.}

\figcaption{Data and the two solid curves same as in Figure 1. 
The dot-dashed curve shows the evolution of [Fe/H] 
at $t>t_{\rm IC}$ for a system in which
infall starts at the big bang but ceases
at $t_{\rm IC}=1.5$ Gyr ($z_{\rm IC}=4$) and the dashed curve
is for $t_{\rm IC}=4.3$ Gyr ($z_{\rm IC}=1.5$). 
The overall evolution of [Fe/H] for the system with 
$t_{\rm IC}=1.5$ Gyr
is indicated by the arrows. In all cases,
star formation starts at the big bang ($t_0=0$). Points
A, B, and C at $z=3$ represent [Fe/H] values sampled by DLA systems
with different $t_{\rm IC}$ values. Assuming
a halo mass of $M_{\rm IC}=10^{11}\,M_\odot$ ($\approx 10^{10}\,M_\odot$ 
in baryonic matter) at the time of infall cessation, $z_{\rm IC}=4$ and 
1.5 correspond to $1.9\sigma$ and $0.95\sigma$ density fluctuations, 
respectively.}

\newpage
\plotone{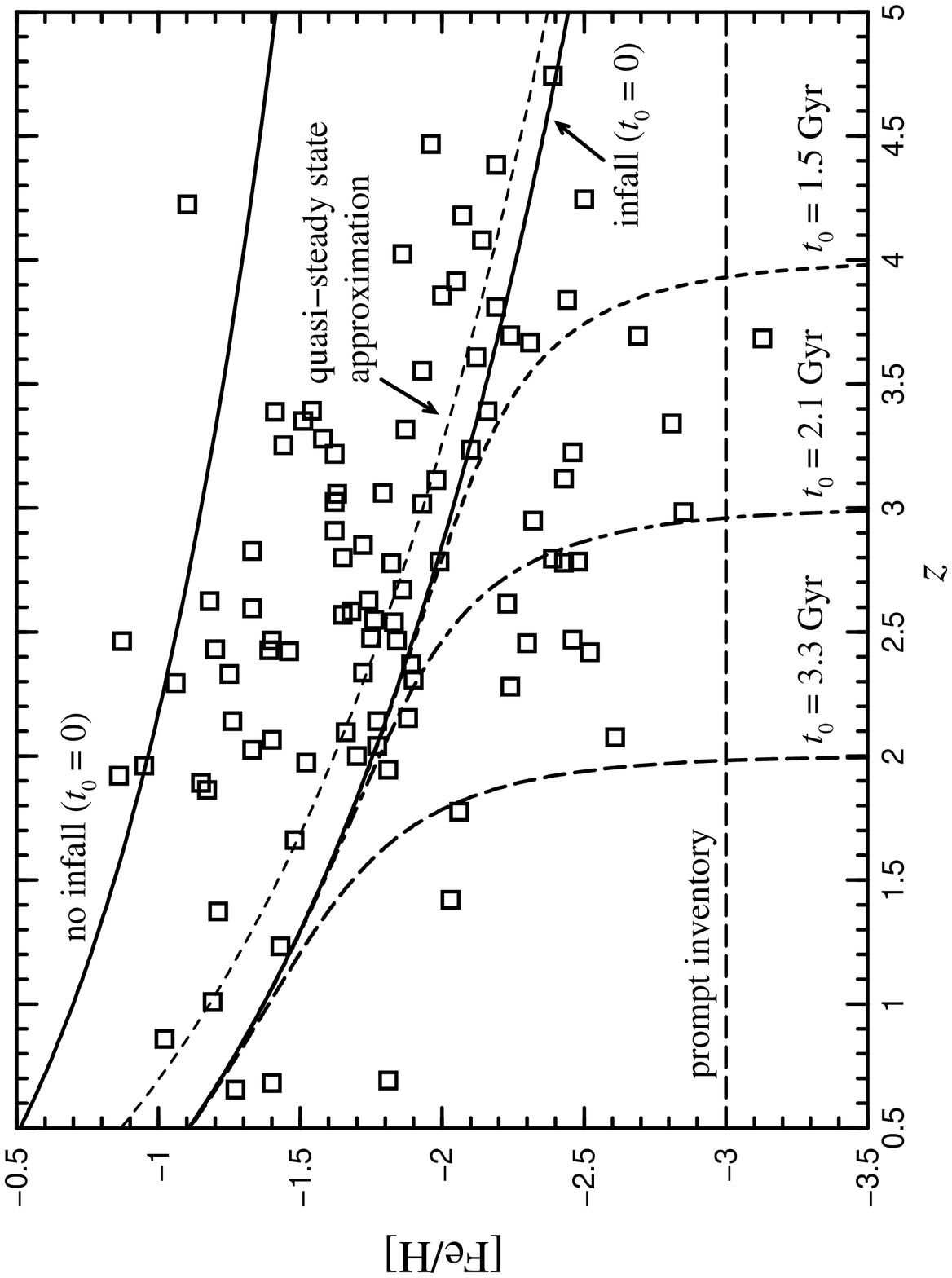}
\newpage
\plotone{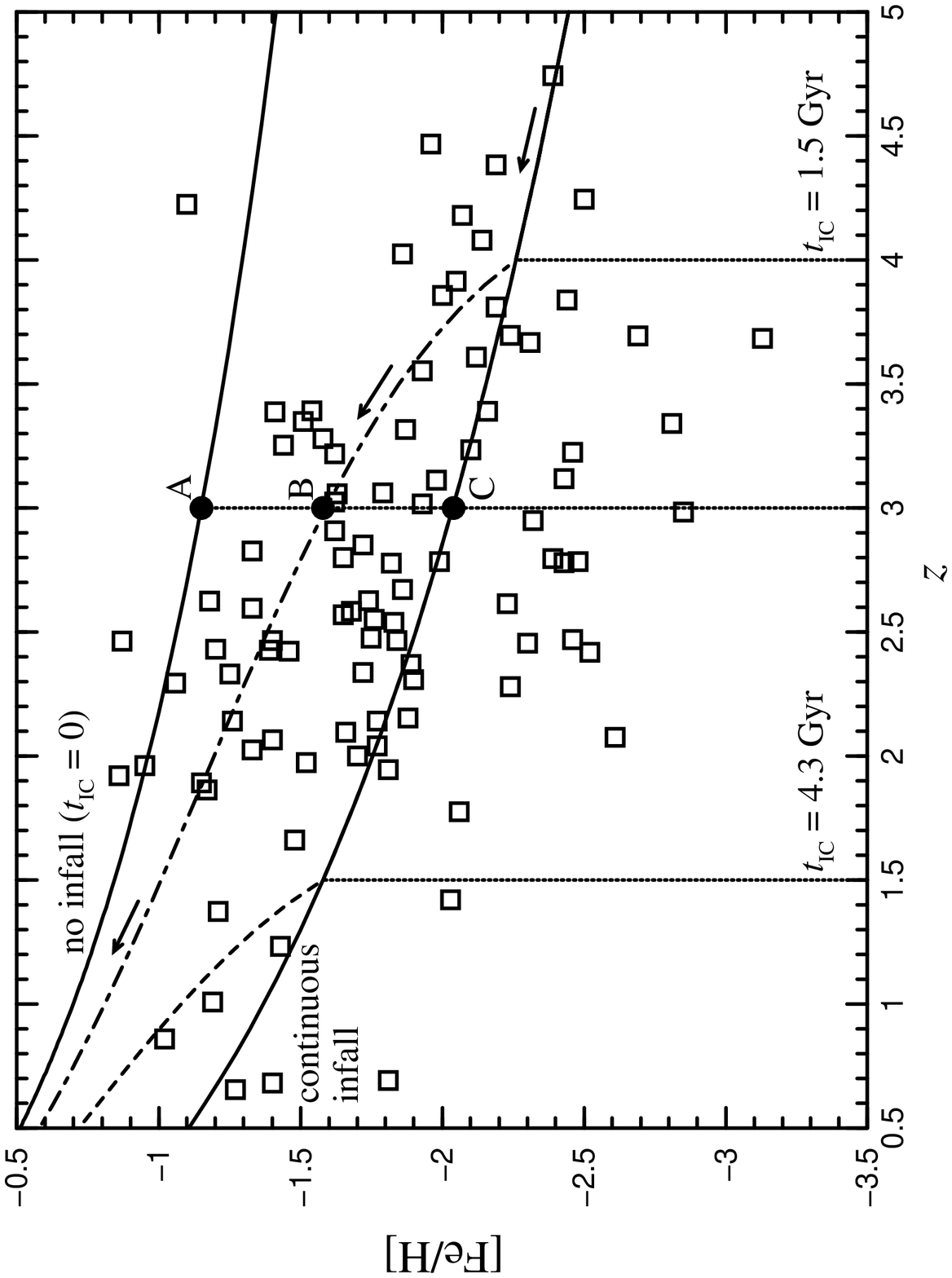}

\end{document}